\documentclass[
superscriptaddress,
final,
showkeys,
pre,
aps,
floatfix,
]{revtex4-2}

\usepackage[dvips]{graphicx}
\usepackage[utf8]{inputenc}
\usepackage{amsmath,amssymb,booktabs,latexsym,MnSymbol,bm}

\usepackage[%
  colorlinks=true,
  urlcolor=blue,
  linkcolor=blue,
  citecolor=blue
]{hyperref}

\begin{document}

\title{Bipartite structure and dynamics of political corruption networks}

\author{M\^onica V. Prates} 
\author{Arthur A. B. Pessa} 
\affiliation{Departamento de F\'isica, Universidade Estadual de Maring\'a, Maring\'a, PR 87020-900, Brazil}

\author{Sebastian~Gon\c{c}alves} 
\affiliation{Instituto de F\'isica, Universidade Federal do Rio Grande do Sul, Porto Alegre, RS 91501-970, Brazil}

\author{Matja{\v z} Perc} 
\affiliation{Faculty of Natural Sciences and Mathematics, University of Maribor, Koro{\v s}ka cesta 160, 2000 Maribor, Slovenia}
\affiliation{Community Healthcare Center Dr. Adolf Drolc Maribor, Ulica talcev 9, 2000 Maribor, Slovenia}
\affiliation{Department of Physics, Kyung Hee University, Dongdaemun-gu, Seoul 02447, Republic of Korea}
\affiliation{Complexity Science Hub, Metternichgasse 8, 1030 Vienna, Austria}
\affiliation{University College, Korea University, Seongbuk-gu, Seoul 02841, Republic of Korea}

\author{Haroldo V. Ribeiro} 
\email{hvribeiro@uem.br}
\affiliation{Departamento de F\'isica, Universidade Estadual de Maring\'a, Maring\'a, PR 87020-900, Brazil}

\date{\today}

\begin{abstract}
Political corruption is inherently an affiliation process linking agents to corruption cases; yet it is often studied via one-mode projections that connect co-offenders within the same scandal, implying a loss of information that potentially confounds properties of agents and cases. Here, we adopt a bipartite representation to analyze datasets of corruption scandals in Brazil and Spain spanning nearly three decades. By tracking the temporal growth of these networks, we quantify density and redundancy measures to capture partner reuse and co-occurrence across cases. Networks in both countries become progressively sparser over time, and agent redundancy is systematically higher than case redundancy, indicating a small cadre of recidivists who recombine largely with novice partners rather than forming durable co-offending ties. These networks exhibit near-exponential degree distributions, reflecting low recidivism and likely high coordination costs and secrecy constraints of large-scale scandals. Our bipartite view further reveals a moderate cross-mode disassortative degree mixing between agents and cases, with high-degree agents distributing their activity across small cases and large scandals mainly comprising low-degree participants. Finally, identifying atypical individuals within the bipartite structure reveals criminal trajectories marked by a gradual rise in network embeddedness that can appear ordinary in agent-projected networks.
\end{abstract}


\maketitle

\section*{Introduction}

Organized crime is not a collection of isolated acts; instead, sophisticated partnerships, division of labor, logistics routines, and secrecy turn otherwise local acts into coordinated criminal enterprises. These features make organized crime a paradigmatic example of complex systems, in which collective patterns underpin the success, adaptation, and resilience of criminal activities. The often-depicted and somewhat cartoonish scene of investigators scrutinizing a wall of suspect photographs linked by strings indeed reflects the core scientific premise that organized crime is relational, and the structure of ties matters. Concepts and tools from network science have therefore become central to the study of organized crime, offering statistical, structural, and modeling lenses on illicit association~\cite{d2015statistical, bouchard2020collaboration, kertesz2021complexity, luna2020corruption, granados2021corruption}. 

Proof of that is a growing body of network analyses spanning mafia groups~\cite{varese2013structure, cavallaro2020disrupting}, illicit drugs~\cite{zakimi2023sociometric}, resilience of drug trafficking~\cite{duijn2014relative}, modular structure of crime organizations~\cite{calderoni2017communities}, political corruption networks~\cite{ribeiro2018thedynamical, diviak2019structure, martins2022universality}, police criminal intelligence networks~\cite{da2018topology, toledo2023multiplex, toledo2025outlier}, cartels in public auction markets~\cite{wachs2019network}, identification of corrupt politicians via voting networks~\cite{colliri2019analyzing}, money laundering~\cite{garcia2020ai}, dark web pedophile rings~\cite{dacunha2020assessing, divakarmurthy2024unravelling}, controllability of criminal networks~\cite{solimine2020political}, dark web marketplaces~\cite{elbahrawy2020collective}, criminal conspiracy networks of shell companies~\cite{nicolas2021conspiracy, nicolas2024organized}, organized crime during the Prohibition era in the United States~\cite{smith2020exogenous, joseph2021ties}, corruption risk in contracting markets~\cite{wachs2021corruption}, graph representation learning to predict properties of criminal networks~\cite{lopes2022machine, ribeiro2023deep}, maritime criminal networks~\cite{chiang2024blend}, and the role of gender in political corruption~\cite{pessa2025structural}. Within this literature, political corruption occupies a distinctive place. Unlike many profit-seeking criminal enterprises that operate largely outside the state, political corruption is embedded in public institutions and party structures, with illicit exchanges routed through procurement, regulatory forbearance, appointments, and legislative favors, yielding wide-ranging societal harms, including reduced economic growth~\cite{rose1975economics, shleifer1993corruption, mauro1995corruption, bardhan1997corruption, shao2007quantitative}, diminished returns of public investments~\cite{haque2008public}, and increased socioeconomic inequality~\cite{mauro1995corruption, gupta2002does}. 

Empirically, information on political corruption typically becomes public through scandals investigated by public prosecutors, parliamentary inquiries, or journalistic expos\'es that assemble elected officials, bureaucrats, party financiers, brokers, and private contractors into coalitions that ultimately convert political influence and public decisions into illicit private gains. Previous studies on political corruption reconstruct these criminal networks by linking agents who co-offended in the same corruption scandal. This one-mode representation has yielded important findings about growth patterns, modular organization, and predictive features of corruption networks~\cite{ribeiro2018thedynamical, martins2022universality, lopes2022machine, ribeiro2023deep}. However, political corruption is inherently an affiliation process linking two distinct types of entities: agents and cases. Ignoring this bipartite nature implies a loss of information~\cite{latapy2008basic, wang2024bipartite, neal2024pattern} that may confound properties of agents and scandals and hide the identification of atypical behavior. Outlier identification in projected networks tends to privilege agents' centrality and overlooks the mode-specific structure that distinguishes, for example, highly active agents involved in many small corruption cases from agents with many partners in a single large scandal. 

Building on prior work with Brazilian and Spanish corruption datasets~\cite{ribeiro2018thedynamical, martins2022universality}, we adopt a bipartite perspective to analyze these criminal networks in their native two-mode form by treating agents and scandals symmetrically, preserving case-level heterogeneity, and enabling separate diagnostics to quantify how agents combine partners across cases and how scandals assemble participants. In doing so, our research investigates network properties that have no natural analogue in one-mode projections, such as mode-specific redundancy and agent-case degree mixing, revealing cross-mode regularities that single-node projections obscure. This bipartite view also allows us to disentangle three fundamental dimensions of agent participation: number of cases, total number of partners, and typical case size. Mining this feature space with an unsupervised anomaly detection algorithm further yields agent-level scores that can be tracked across time, enabling longitudinal profiles of atypical involvement that do not always coincide with central positions in the agent projections.

In what follows, we first describe our datasets of well-documented political scandals in Brazil and Spain from the late 1980s to the late 2010s. We then introduce the bipartite representation and analyze the temporal evolution of topological properties. Next, we present an approach to identify atypical agents and contrast their profiles with properties obtained from the agent projection. Finally, we conclude by summarizing our findings and outlining their implications.

\section*{Data}

Our investigation relies on previously compiled datasets of political corruption scandals from Brazil~\cite{ribeiro2018thedynamical} and Spain~\cite{martins2022universality}, each listing the names of individuals involved in every scandal and the year the cases became public. The Brazilian dataset comprises 65 scandals between 1987 and 2014, involving 404 distinct agents, compiled from well-documented cases initially assembled in the Wikipedia list~\cite{wikilist} and cross-referenced with national newspapers and magazines. The Spanish dataset comprises 437 scandals between 1989 and 2018, involving 2,695 distinct agents, and builds on the non-profit \emph{Casos Aislados} initiative~\cite{casosaislados}, which aggregates scandals reported by the Spanish press. As emphasized in the original data sources~\cite{ribeiro2018thedynamical, martins2022universality}, data on illicit activity obtained from media vehicles are subject to limitations, including incomplete coverage, as some participants may evade detection, and reporting bias, as high-profile scandals may receive disproportionate media attention. Official records of corruption investigations -- which themselves are not free from bias or incompleteness -- are often restricted by legal constraints and typically accessible only to the public prosecutor's office and the judiciary~\cite{bright2022reprint}. Nevertheless, our analyses focus on structural regularities that consistently emerge across two independently curated national corpora with distinct institutional contexts -- although limitations certainly exist, they are more likely to attenuate than materially affect the results we report.

\section*{Results}

\begin{figure*}[ht!]
    \centering
    \includegraphics[width=0.9\linewidth]{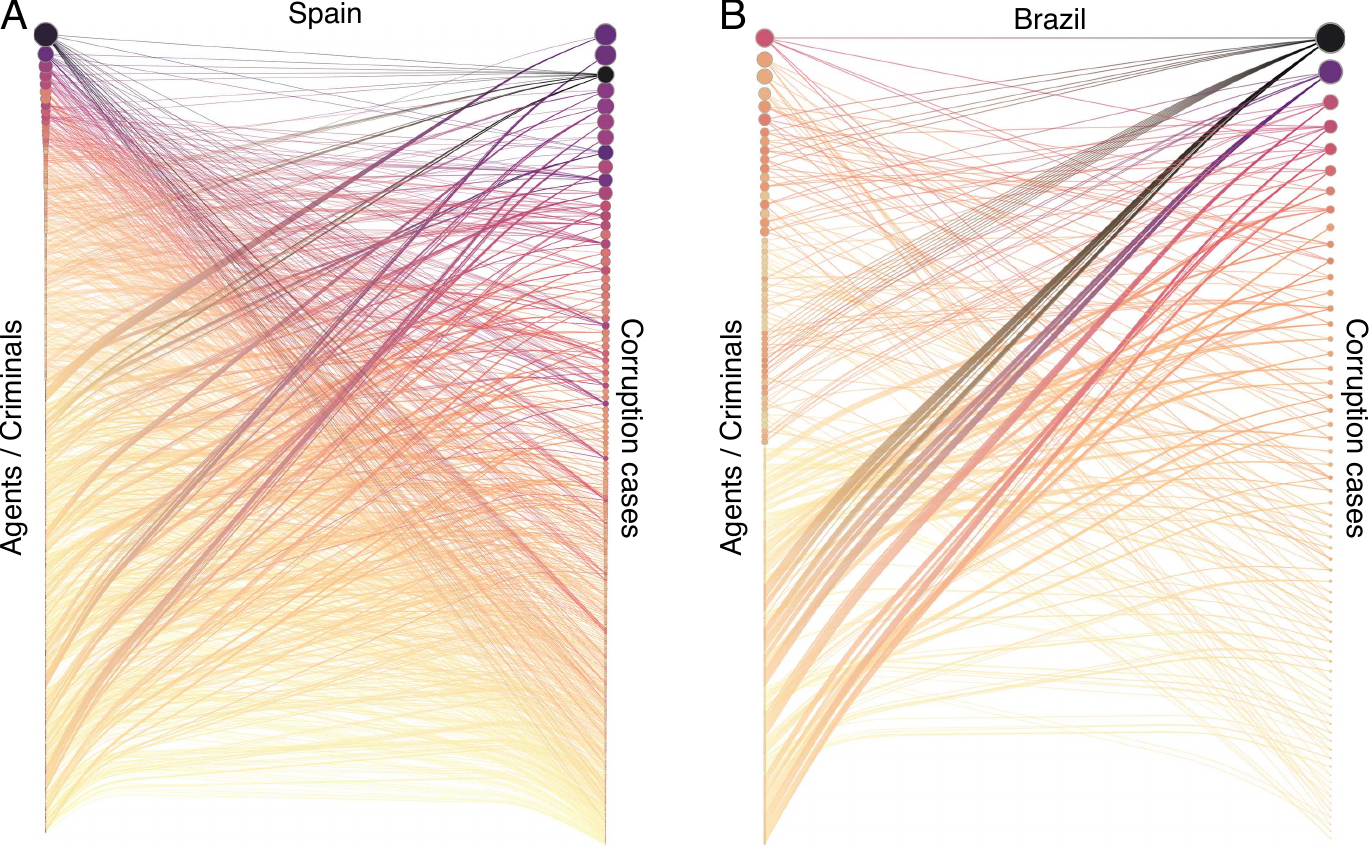}
    \caption{Bipartite structure of corruption networks. Visualizations of the bipartite corruption networks for (A) Spain and (B) Brazil. In each panel, nodes on the left represent individuals and nodes on the right represent corruption scandals. Edges link individuals to the scandals in which they participated. Nodes are ordered from top to bottom by descending degree and are colored by score defined as the sum, over all incident edges, of the product of the degrees of the two endpoints; edge colors interpolate between the colors of their endpoint nodes (darker shades indicate larger values). These visualizations highlight that agents involved in many cases tend to distribute their participation across small and large cases, whereas agents involved in few cases are predominantly associated with the largest corruption scandals, yielding a structure marked by high-degree individuals involved in multiple cases and large scandals encompassing many low-degree participants.}
    \label{fig:1}
\end{figure*}

We begin by constructing bipartite networks that encompass all corruption scandals documented in our datasets for Spain and Brazil. Figure~\ref{fig:1} depicts a visualization of the Spanish (panel A) and Brazilian (panel B) networks. In each panel, nodes on the left represent agents, while nodes on the right correspond to the political corruption cases. Edges connect agents to the cases in which they participated. Nodes of both types are vertically ordered by decreasing degree and color-coded by a node-level score defined as the sum, over all incident edges, of the product of the degrees of the two endpoints (darker shades indicate larger values). Edges are colored using a gradient that interpolates between the colors of their endpoint nodes. These visualizations indicate that agents involved in many cases tend to appear both in large and small cases, whereas agents involved in few cases are predominantly associated with the largest cases. 

\begin{figure*}[ht!]
    \centering
    \includegraphics[width=0.9\linewidth]{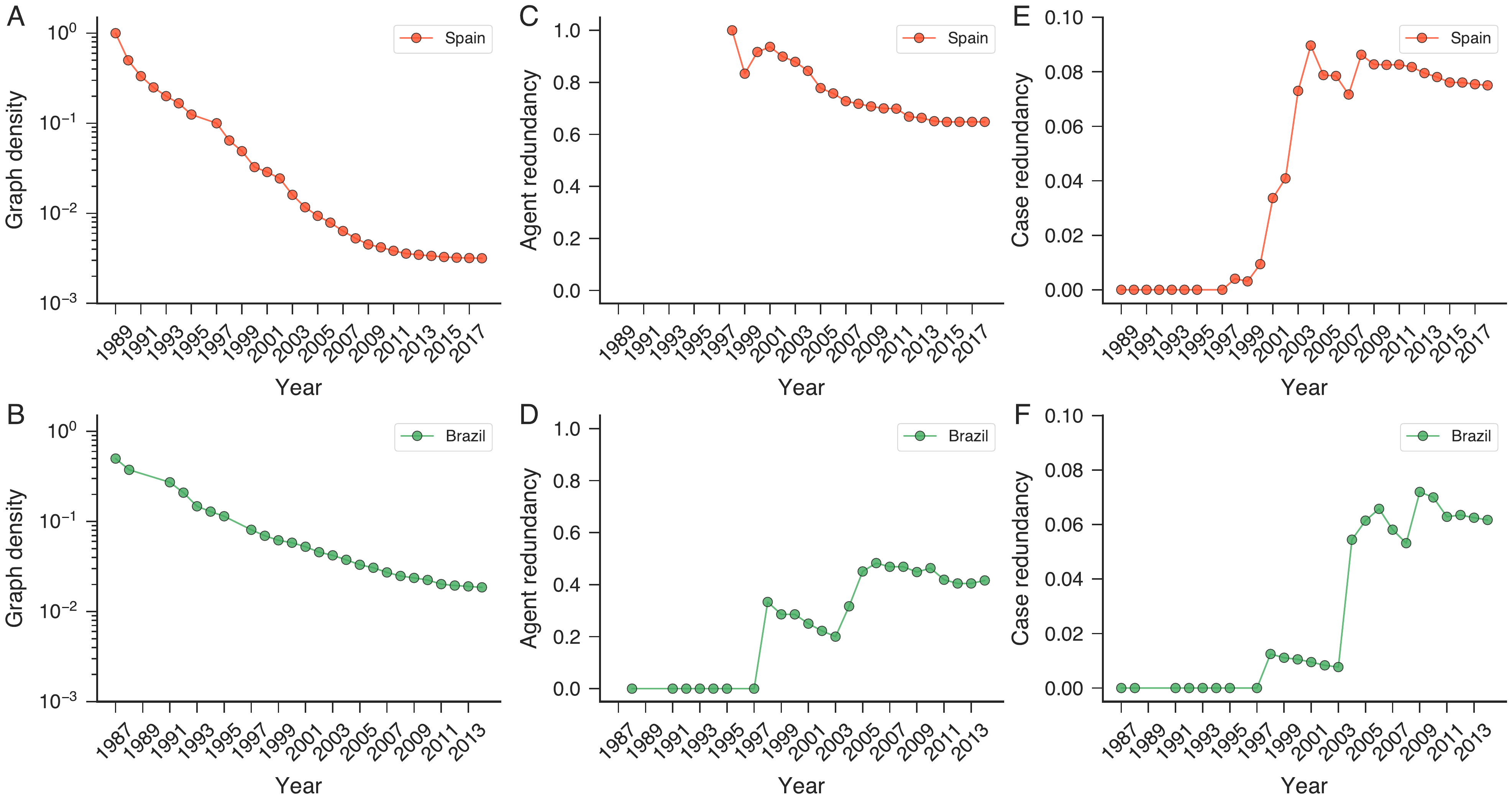}
    \caption{Evolution of topological properties during the growth of bipartite corruption networks. Panels show (A,B) graph density, (C,D) agent redundancy, and (E,F) corruption-case redundancy for the Spanish (top) and Brazilian (bottom) networks. Density is the proportion of realized agent–case edges among all possible. For an agent, redundancy is the fraction of pairs of its case neighbors that are also jointly linked by some other agent; the network-level agent redundancy is the mean over all agents. Case redundancy is defined analogously for each case as the fraction of pairs of its agent neighbors that also co-occur in at least one other case, averaged over cases. The persistent decline in density shows that both networks become progressively sparser over time, indicating that network expansion is primarily driven by the entry of new agents and cases rather than repeated collaborations among known actors, further suggesting that most scandals are short-lived and likely transaction-specific coalitions. Both networks consistently exhibit agent redundancy significantly higher than case redundancy, likely reflecting a small cadre of recidivists who repeatedly mobilize novice collaborators across scandals, consistent with secrecy and risk management strategies that discourage persistent criminal partnerships.}
     \label{fig:2}
\end{figure*}

To start quantifying structural patterns and how they possibly evolve, we reconstruct the networks cumulatively by year, adding agents and scandals as they were discovered. We first calculate, for each year, the network density defined as the proportion of realized agent–case edges among all possible~\cite{latapy2008basic}. Figures~\ref{fig:2}A and \ref{fig:2}B show that density declines as both networks grow, reaching very low values (0.003 for Spain and 0.019 for Brazil) after the inclusion of all scandals. This progressive sparsification indicates that the expansion of these networks is primarily driven by the entry of new agents and cases rather than by repeated collaborations among known actors. This limited reuse of criminal partnerships further suggests that most scandals are short-lived and transaction-specific coalitions. Such ephemeral structuring likely reflects strong concealment incentives, as reusing partners increases traceability through shared communication and financial trails exposed by prior investigations, but also indicates the presence of a small cadre of recidivists who repeatedly assemble new teams composed mostly of first-time offenders unlikely to reappear in other scandals.

We next examine the redundancy coefficient at the agent and case levels~\cite{latapy2008basic}. For an agent, redundancy is the fraction of pairs of its case neighbors that are also jointly linked by another agent. Analogously, case redundancy is defined as the fraction of pairs of a case's agent neighbors that co-occur in at least one other case. Both quantities are defined only for nodes with two or more links, and network-level values are the average values across the redundancy coefficient of such agents. Because triangles cannot occur in bipartite networks, redundancy provides an analogue of local transitivity via four-cycles, capturing how agents recycle subsets of partners across distinct cases and how agent pairs reappear together in multiple cases. Figures~\ref{fig:2}C and \ref{fig:2}D depict the evolution of average agent redundancy for the Spanish and Brazilian networks, respectively, whereas Figures~\ref{fig:2}E and \ref{fig:2}E present the analogous evolution for average case redundancy. For both countries, agent redundancy significantly exceeds case redundancy over the cumulative reconstruction of the criminal networks, with higher final-stage values in Spain for both measures. Spanish agent redundancy, undefined before 1998 due to the absence of agents participating in two or more cases, shows a decline before stabilizing, whereas the Brazilian series exhibits abrupt increases in 1997 and again after 2003 before reaching a plateau. Spanish case redundancy rises sharply around 2000 before plateauing, with a qualitatively similar behavior observed in Brazil around 2003. Consistent with the low densities, these patterns likely reflect secrecy and risk management that discourage persistent criminal partnerships between agents, keeping case redundancy low. In turn, higher agent redundancy is consistent with the existence of a small cadre of recidivists who repeatedly mobilize new collaborators across scandals. 

\begin{figure*}[t!]
    \centering
    \includegraphics[width=0.99\linewidth]{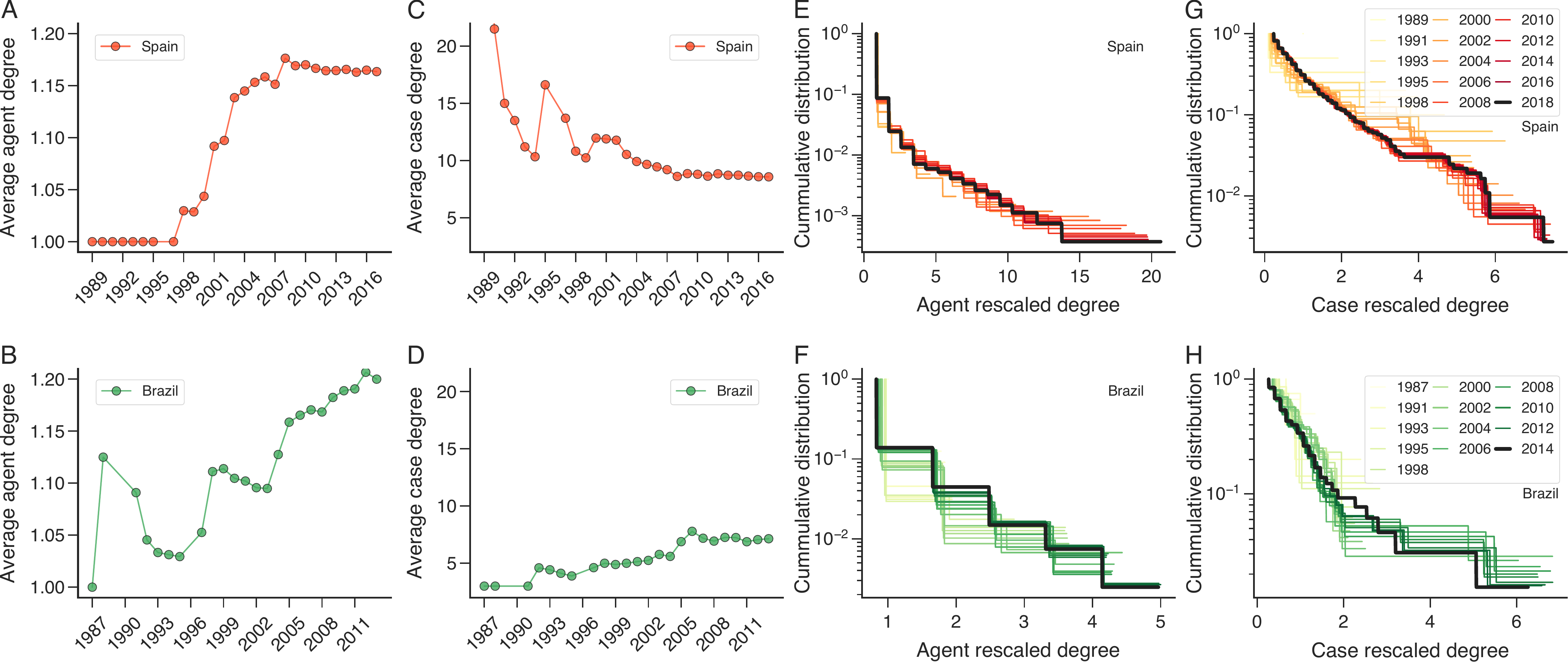}
     \caption{Degree dynamics of agents and cases during the growth of bipartite corruption networks. Evolution of the average degree of agents (number of cases per agent) in the (A) Spanish and (B) Brazilian networks. Evolution of average case degree (number of agents per case) for the (C) Spanish and (D) Brazilian networks. Yearly cumulative distributions of the rescaled agent degree (degree divided by the corresponding yearly average) for (E) Spain and (F) Brazil. Yearly cumulative distributions of the rescaled case degree for (G) Spain and (H) Brazil. Cumulative distributions are color-coded by year, with black curves denoting the last available year. The stabilization of the average agent degree at small values reflects low recidivism among political agents. Similarly, the stabilization of the average case degree at small numbers implies that typical corruption cases are small, a scale likely enforced by coordination costs and secrecy constraints. The rescaled cumulative degree distributions for both agents and cases exhibit consistent collapse across years and present near-exponential forms on semi-log plots, indicating a lack of scale-free behavior. However, the pronounced deviations in the tails of the case degree distributions suggest that the largest scandals are generated by mechanisms distinct from those governing typical cases.}
     \label{fig:3}
\end{figure*}

Turning to degree structure, we characterize the evolution of the average agent and case degrees and their distributions as the networks grow. Figures~\ref{fig:3}A and \ref{fig:3}B show the dynamics of the average agent degree for Spain and Brazil, respectively, whereas Figures~\ref{fig:3}C and \ref{fig:3}D present the corresponding results for the average case degree. In Spain, the average agent degree (cases per agent) remains at one until 1998 -- reflecting the absence of agents appearing in multiple cases -- then rises through 2008 and stabilizes at $1.16$. In Brazil, it increases more steadily (albeit with larger early fluctuations) and reaches $1.21$ in the final network stage. These averages are thus small and similar for both countries, reflecting the limited repeated participation of agents, in which most appear once and only a small minority engages in repeat offending. This finding is in agreement with previous findings~\cite{martins2022universality} and contrasts with violent and property crimes, for which recidivism is much larger~\cite{alper2018update}. In its turn, the average case degree (agents per case) in Spain shows a decreasing trend until 2007 and stabilizes at $8.6$, while in Brazil it grows until 2005 and then saturates around $7.1$. The stabilization of average case degree in small numbers implies that typical corruption episodes are small, a scale that likely reflects coordination costs and secrecy constraints~\cite{ribeiro2018thedynamical, martins2022universality}.

To probe distributional shape beyond the averages, we also analyze the evolution of the cumulative distributions of degrees normalized by their yearly averages, as shown in Figures~\ref{fig:3}E and \ref{fig:3}F for the agent rescaled degree in Spain and Brazil, respectively, and in Figures~\ref{fig:3}G and \ref{fig:3}H for the case rescaled degree. We observe that curves for agent degree from different years collapse well and present an approximately linear behavior on the semi-log plots, implying near-exponential forms. Nonetheless, the Spanish distributions exhibit heavier tails, indicating a higher level of heterogeneity among recidivist agents, which in turn is consistent with a small set of highly recurrent agents in Spain (for instance, one involved in 24 cases), whereas multiple involvement is much more limited in Brazil (the maximum is a single agent involved in six cases). The case-rescaled degree distributions from different years also collapse well and likewise appear approximately linear on semi-log plots, suggesting near-exponential forms and a lack of scale-free behavior. The distributions for both countries, however, exhibit pronounced deviations from this exponential form in the tails, suggesting that the largest scandals are generated by mechanisms distinct from the typical case, where episodic opportunities yield rare and very large criminal coalitions (up to 64 agents in Spain and 47 in Brazil). Moreover, the existence of these atypical cases, combined with the stabilization of the average agent degree, appears to agree with life-course studies of white-collar offenders~\cite{weisburd2001white} reporting heterogeneous, opportunity-driven trajectories marked by intermittent reoffending concentrated in a small subset of criminals whose accumulated access and know-how may enable disproportionate participation in corruption cases. 

\begin{figure*}[ht!]
    \centering
    \includegraphics[width=0.6\linewidth]{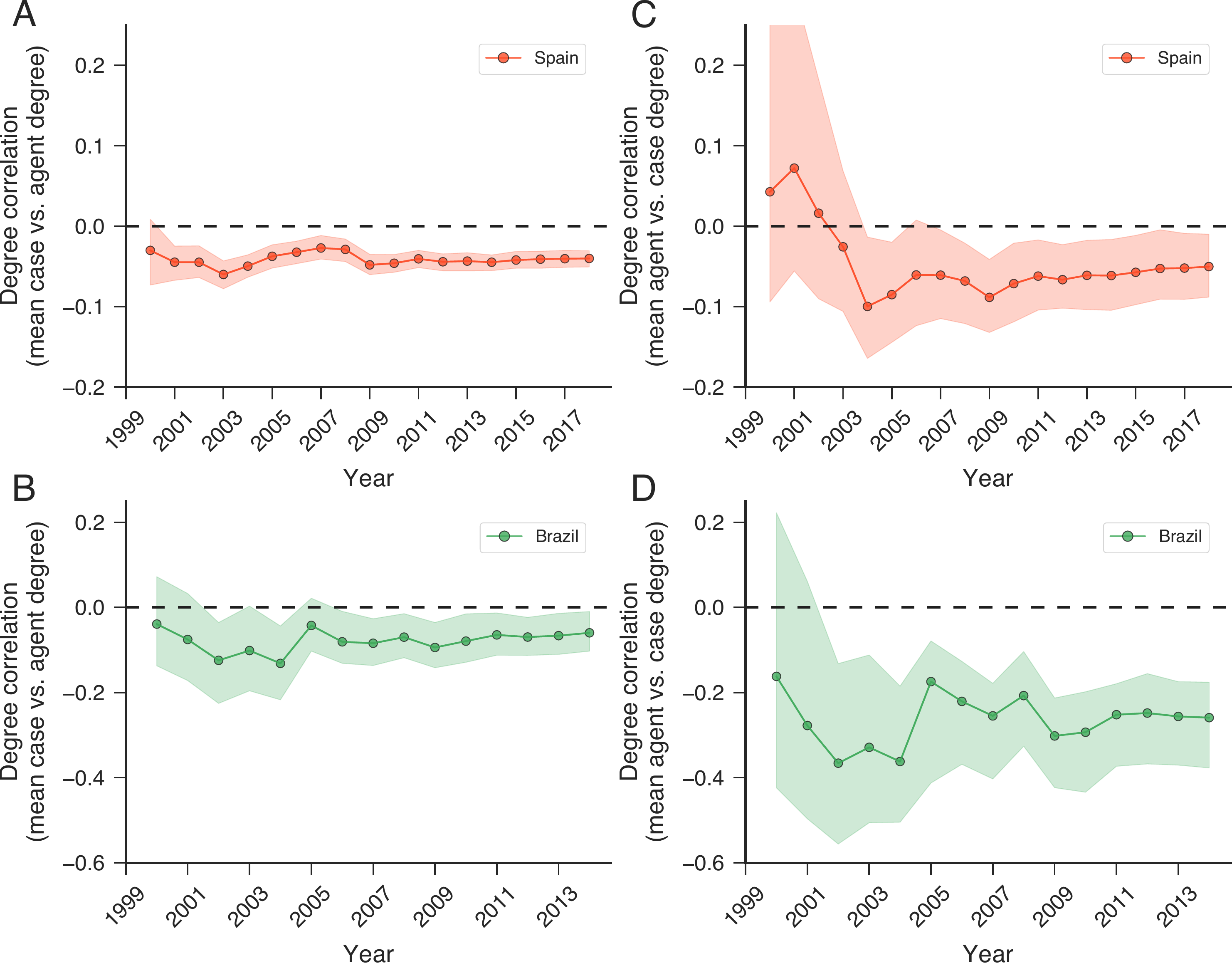}
     \caption{Disassortative degree mixing between agents and corruption cases. Evolution of the Pearson correlation between the average degree of the cases in which each agent appears and their agent degree for the (A) Spanish and (B) Brazilian networks. Evolution of the Pearson correlation between the average degree of agents participating in each case and the case degree for the (C) Spanish and (D) Brazilian networks. Shaded bands show 95\% bootstrap confidence intervals and horizontal dashed lines denote zero correlation. The small yet persistent negative correlations across most years indicate a moderate cross-mode disassortative mixing between agents and cases. This means that high-degree agents (those involved in many cases) are not concentrated in the largest scandals but tend to appear in smaller cases, whereas large scandals are populated mainly by low-degree participants. This pattern is consistent with secrecy and coordination constraints, as well as risk-spreading strategies employed by recidivists, who likely avoid co-appearing with other experienced agents in large coalitions. Structurally, this disassortativity suppresses hub–hub coalescence, favors many small scandals loosely stitched together, and delays core densification.}
     \label{fig:4}
\end{figure*}

Our bipartite framework further enables us to relate each agent's activity to the typical size of the cases they join -- and, symmetrically, to relate each case's size to the average activity of its participants -- relationships that one-mode projections can obscure~\cite{latapy2008basic}. We quantify these associations by tracking, over network growth, the Pearson correlation between the average degree of the cases in which each agent appears and their agent degree for Spain and Brazil, as shown in Figures~\ref{fig:4}A and \ref{fig:4}B, respectively. Symmetrically, we calculate the correlation between the average degree of agents participating in each case and the case degree for Spain and Brazil, as shown in Figures~\ref{fig:4}C and \ref{fig:4}D, respectively. In both countries and for both node types, correlations are small yet statistically significant and persistently negative across most years. This pattern indicates a moderate cross-mode disassortative mixing between agents and corruption cases, in which high-degree agents tend to distribute their participation across smaller cases, while large scandals are populated mainly by low-degree participants. Such disassortativity is consistent with secrecy and coordination constraints, as well as risk-spreading strategies employed by recidivists, who likely avoid co-appearing with other experienced agents in large coalitions. It also helps suppress hub–hub coalescence, favors many small scandals loosely stitched together by thin bridges, and delays core densification and giant-component consolidation -- findings that agree with the fact that agent-projected corruption networks appear to operate near a critical recidivism rate, below which they become entirely fragmented, whereas above it they become overly connected~\cite{martins2022universality}.

\begin{figure*}[t!]
    \centering
    \includegraphics[width=0.9\linewidth]{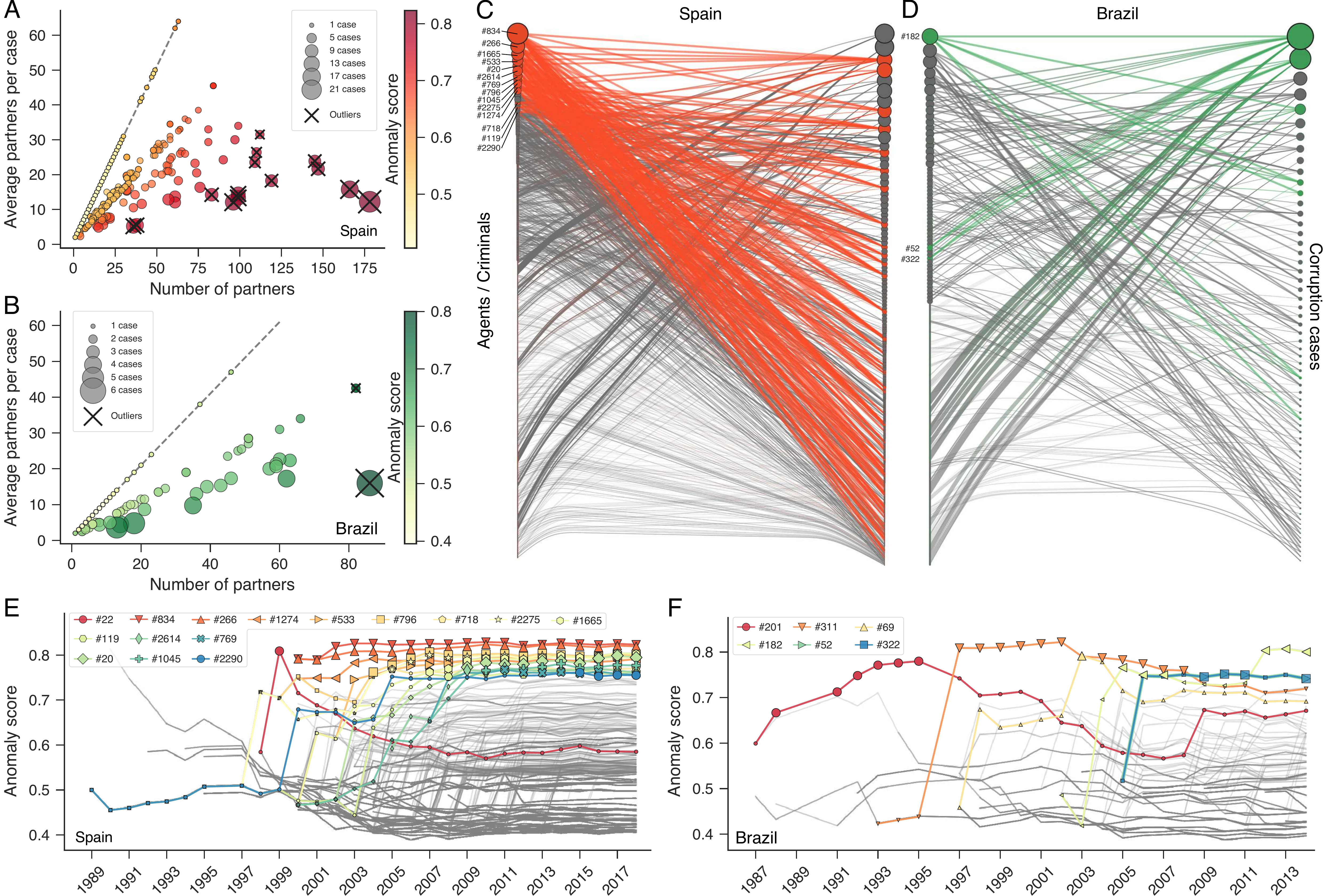}
     \caption{Mining outliers in bipartite corruption networks. Scatter plot depicting agents' number of partners, average number of partners per case, and number of corruption cases for the (A) Spanish and (B) Brazilian networks. In both previous panels, results correspond to the final stage of each network. Dashed lines show the 1:1 relationship between the average number of partners per case and the total number of partners. Agents on this line participate in a single case. Cross markers indicate outliers identified with the isolation forest algorithm, and points are color-coded by their anomaly scores. The wide distribution across the three dimensions corroborates the need for a bipartite framework to disentangle recidivism, collaborative breadth, and exposure to large scandals. Visualization of agents flagged as outliers in (C) Spain and (D) Brazil. Outlier agents, their incident edges, and the cases they join are highlighted in color, while all remaining nodes and edges are in gray. This visualization confirms that atypical agents are often involved in small- and mid-sized scandals rather than being concentrated in the largest cases, reducing their average number of partners per case. Evolution of the anomaly score for each agent in the (E) Spanish and (F) Brazilian networks. Agents identified as outliers in at least one year are labeled and displayed in color, with larger markers indicating years in which they were classified as outliers. The final stage of the Brazilian network contains three outliers; two of which share identical properties and therefore overlap in panels (B) and (F). The temporal analysis further reveals that the criminal trajectories of most atypical agents are marked by a gradual ascent in the anomaly-score ranking over several years before reaching outlier status, suggesting that the emergence of atypical involvement is often a career process involving an accumulation of opportunities and know-how, rather than an isolated sequence of events.}
     \label{fig:5}
\end{figure*}

In light of these structural regularities, a practical concern is that central agents may be masked in the agent-projected network. Indeed, large cases induce dense cliques that inflate co-occurrence degrees for many individuals who have low degrees in the bipartite networks. In contrast, highly active criminals operating across numerous small cases accrue relatively few pairwise ties per case and may appear only moderately central under standard centrality measures evaluated from agent-projected networks. Our bipartite representation is thus essential for detecting outlier individuals, as it allows us to disentangle three fundamental dimensions of agent participation: number of cases, total number of distinct partners, and typical case size (the mean degree of the cases an agent joins). This triad decouples recidivism, collaborative breadth, and exposure to large scandals, enabling the identification of agents whose participation patterns are atypical relative to the population baseline. Figures~\ref{fig:5}A and \ref{fig:5}B show scatter plots of these three dimensions for every agent in the final stages of the Spanish and Brazilian networks, respectively. In this visualization, the $x$-axis gives the total number of distinct partners (the agent's degree in the agent-projected network), the $y$-axis the average number of partners across the cases in which each agent was involved (the average degree of the agent's incident case nodes in the bipartite network), and marker size encodes the number of cases per agent (the agent's degree in the bipartite network). Agents are broadly distributed across the three dimensions, and those involved in a single case lie on the 1:1 dashed line. Moreover, consistent with the network properties previously described, agents involved in multiple cases and accumulating many partners often display intermediate values of average partners per case.

Despite lacking a ground truth for comparison, we propose to identify atypical individuals by applying the isolation forest algorithm~\cite{liu2008isolation, liu2012isolation} to our three dimensions of agent participation. While there are many other possible approaches, we consider this method due to its simplicity, minimal number of hyperparameters, low computational cost, interpretability through an anomaly score, and its demonstrated high accuracy in benchmark anomaly detection tasks~\cite{liu2008isolation, liu2012isolation}. Further support comes from recent work on real-world criminal-network datasets, where isolation forest (compared against several unsupervised detectors) scaled efficiently and yielded informative anomaly rankings for disruption analyses~\cite{toledo2025outlier}. Operationally, the algorithm builds an ensemble of random partitioning trees that repeatedly select one of the three variables -- number of cases, total number of distinct partners, and typical case size -- and split points within their ranges to progressively isolate each agent. Individuals that fall in sparse or unexpected regions of this three-dimensional space are isolated with fewer splits than those following common participation patterns. An anomaly score, ranging from $0$ to $1$~\cite{liu2008isolation, liu2012isolation}, is then calculated based on the average number of splits required to isolate an agent across the ensemble, so that higher values indicate easier isolation (more anomalous) and values near or below $0.5$ indicate typical behavior.

In our analysis, we fix the number of trees at $500$ (a value that yields stable scores) and set the expected anomaly proportion (the contamination parameter) at $0.005$. The contamination parameter controls the share of agents flagged as anomalous, and a value of 0.5\% intentionally yields a small number of atypical individuals. These settings are indeed \textit{ad hoc}, and if ground-truth labels were available, hyperparameters could be optimized, for example, by calibrating precision–recall trade-offs via cross-validation. In Figures~\ref{fig:5}A and \ref{fig:5}B, agents flagged as outliers at the 0.5\% level are indicated with a cross; however, anomaly scores for all agents are shown on a continuous color scale for threshold-independent exploration. Fourteen atypical individuals are identified in the final stage of the Spanish network, whereas three outliers -- two sharing identical properties --  emerge in the Brazilian network. These individuals and the cases in which they were involved are further highlighted in the network visualizations depicted in Figures~\ref{fig:5}C and \ref{fig:5}D, where nodes of both types are vertically ordered by decreasing degree. We observe that atypical agents are not necessarily involved in the largest cases -- particularly in Spain, where the two largest cases contain no outliers. Instead, they are predominantly involved in small and mid-sized scandals, which, in turn, reduces their average number of partners per case.

Beyond this static scenery focused on the final stages of our corruption networks, we examine how atypical agents emerge as the networks grow. To do so, we track the three dimensions of agent participation over time (number of cases, total number of distinct partners, and typical case size) and, for each network snapshot, apply the isolation forest algorithm (using the same previous settings) to obtain a time series of anomaly scores starting from each agent's first involvement in a corruption scandal. Figures~\ref{fig:5}E and \ref{fig:5}F show these trajectories for all agents in the Spanish and Brazilian networks, respectively. Grey curves represent agents that were never flagged as outliers, many of which share identical scores and therefore overlap, with opacity reflecting the multiplicity of these agents. In turn, colored curves with distinct markers indicate agents flagged as outliers in at least one snapshot, with larger markers denoting the years in which they were classified as outliers. Some agents become outliers after a few years of their first appearance, and a minority do so from the outset. However, most atypical agents ascend the anomaly-score ranking for several years before reaching the outlier status, as illustrated by the notable case of agent \#2290, who entered at the inception of the Spanish network and became an outlier 27 years later. Moreover, despite the commutative nature of our network reconstruction, we observe a few agents that gain and later lose outlier status as the networks expand, as in the cases of Spanish agent \#22 and Brazilian agent \#69. 

\begin{figure*}[t!]
    \centering
    \includegraphics[width=0.7\linewidth]{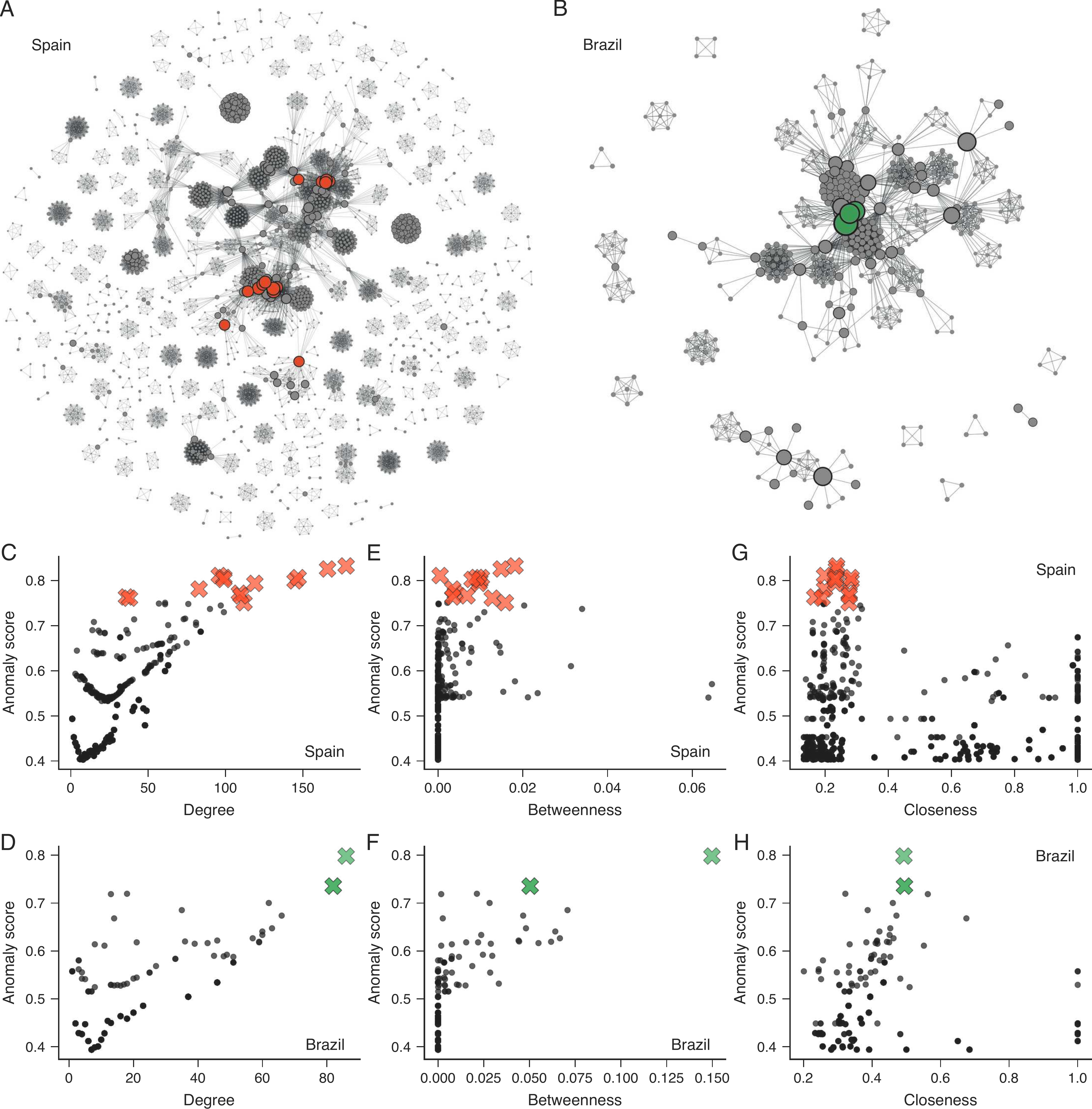}
     \caption{Anomaly score of agents obtained from the bipartite structure of corruption networks compared with simple centrality measures calculated from the agent-projected network. Visualization of agents' projection for the final stages of the (A) Spanish and (B) Brazilian networks. Nodes represent agents and edges link agents who were accomplices in at least one corruption case. Node size is proportional to degree, and colored nodes denote outliers. These visualizations confirm that some outlier agents appear ordinary or even peripheral in the agent-projected networks. Relationship between the agents' anomaly scores and their (C,D) degrees, (E,F) betweenness centralities, and (G,H) closeness centralities for the Spanish (C,E,G) and Brazilian (D,F,H) networks. Results correspond to the final stage of each network, and colored crosses indicate outliers. Two of the three outliers in the Brazilian network share the same parameters and therefore overlapped in the panels. These comparisons reveal that, although degree centrality correlates moderately with the anomaly score, betweenness and closeness centralities exhibit much weaker associations, confirming that agents identified as atypical using the bipartite structure do not necessarily occupy the most central positions in the one-mode projections.}
     \label{fig:6}
\end{figure*}

Finally, we compare the anomaly scores of agents obtained from the bipartite networks with standard centrality measures computed on the agent-projected networks. Figures~\ref{fig:6}A and \ref{fig:6}B illustrate the one-mode projections for the Spanish and Brazilian networks, respectively, where nodes represent agents and edges indicate co-participation in at least one corruption scandal. These projections aggregate all corruption cases in each country, with atypical individuals identified from the bipartite analysis highlighted by colored circles. As argued before, some outlier agents appear ordinary or even peripheral in these agent-projected networks. To corroborate this observation, we calculate simple centrality measures on the agent-projected networks and compare them with the anomaly scores. Figures~\ref{fig:6}C and \ref{fig:6}D show anomaly score versus degree for the Spanish and Brazilian networks, respectively; Figures~\ref{fig:6}E and \ref{fig:6}F do the same for betweenness centrality; and Figure~\ref{fig:6}G and \ref{fig:6}H for closeness centrality. In all scatter plots, outlier agents are indicated by colored cross markers. Degree centrality correlates moderately with the anomaly score, whereas betweenness and closeness show much weaker associations. These results thus demonstrate that agents identified as outliers in the bipartite representation do not necessarily occupy the most central positions in the one-mode projections; even for degree, where the association is stronger, several Spanish outliers do not exhibit the largest centrality values.

\section*{Conclusions}

We thus modeled political corruption as a two-mode affiliation system and showed that this bipartite lens reveals regularities that one-mode projections obscure. Using datasets on political corruption scandals from Spain and Brazil spanning nearly three decades, we found that bipartite corruption networks become progressively sparser as they grow, indicating an expansion mechanism primarily driven by the entry of new agents and cases rather than by increasing the number of co-offending ties among known agents. Our bipartite view further allowed us to quantify how agents rely on subsets of partners across cases and how agent pairs appear in multiple cases through measures of agent and case redundancies. Results demonstrated that agent redundancy consistently exceeds case redundancy, which in turn suggests the existence of a small cadre of recidivists who recombine largely with novice partners across scandals and avoid forming more durable co-offending ties. We also observed that these networks exhibit near-exponential degree distributions, with more pronounced tail deviations in case degrees, suggesting that large criminal coalitions are likely driven by distinct mechanisms from those ruling typical-sized scandals. Moreover, the stabilization of the average case degree at small values in both countries appears consistent with coordination costs and secrecy constraints~\cite{ribeiro2018thedynamical, martins2022universality}.

Our framework further allowed us to quantify the relation between agents' activity and the typical size of corruption cases they joined, as well as between case size and the average activity of its agents -- associations with no direct analogy in one-model projections. We found moderate cross-mode disassortative degree mixing between agents and cases, indicating that high-degree agents distribute their activity across small cases and large scandals mainly comprise low-degree participants. Combined with previous structural regularities, this pattern suggests that standard centrality measures evaluated in agent-projected networks may fail to reveal highly active criminals operating across numerous and predominantly small corruption cases. We therefore identified atypical individuals through three fundamental dimensions of agent participation embedded in the bipartite structure: number of cases, total number of distinct partners, and typical case size. Using this representation space with an anomaly detection algorithm, we confirmed that some atypical individuals appear ordinary or even peripheral in agent-projected networks, as measured by degree, betweenness, and closeness centralities. By tracking the evolution of an anomaly score since the first involvement of each agent, we further observed that some agents become outliers just a few years after their first appearance. Yet, many others display long ascending trajectories in the anomaly-score ranking before reaching outlier status, and only a minority gain and later lose outlier status as the networks grow. Criminal trajectories for atypical individuals thus appear as a career process rather than a sequence of isolated events. The gradual rise of most atypical agents suggests the existence of an incubation period during which individuals accumulate opportunities, assess risks, build trust, and secure access that may favor subsequent criminal activities -- patterns that seem consistent with the social learning theory~\cite{akers2009social}, which frames criminal conduct as a behavior learned through peer association, vicarious learning, and rewards that outweigh sanctions.

Our work does not go without its limitations. Media-compiled corpora are inevitably incomplete and susceptible to reporting biases, and the absence of ground-truth labels for atypical agents precludes external validation of anomaly classifications. These limitations are, however, intrinsic to such illegal activities, and even official records, which rarely become public due to legal constraints, are not free from bias or incompleteness. Even so, the emergence of similar structural regularities across two independently curated national datasets from distinct institutional contexts provides robustness and substantive value to our findings. We further highlight that the methodological approach used here is readily applicable to studying complex affiliation systems across various domains beyond political corruption. This bipartite perspective is valuable for analyzing other organized crime structures, such as criminal conspiracy networks involving shell companies, dark web marketplaces, maritime criminal networks, drug trafficking organizations, and mafia groups. By modeling the relationships between individuals and the operations, ventures, or enterprises they participate in, we preserve mode-specific structural information that traditional one-mode projections obscure. Moreover, our outlier identification approach based on the three fundamental dimensions of agent participation (number of cases, total number of distinct partners, and typical case size) is entirely transferable to any bipartite dataset where detecting statistically atypical individuals is important

\section*{CRediT authorship contribution statement}
\noindent {\bf M\^onica V. Prates:} Conceptualization, Methodology, Software, Investigation, Data Curation, Visualization, Writing - Original Draft, Writing - Review \& Editing. {\bf Arthur A. B. Pessa:} Conceptualization, Methodology, Software, Investigation, Data Curation, Visualization, Writing - Original Draft, Writing - Review \& Editing. {\bf Sebastian~Gon\c{c}alves:} Conceptualization, Methodology, Software, Investigation, Data Curation, Visualization, Writing - Original Draft, Writing - Review \& Editing. {\bf Matja{\v z} Perc:} Conceptualization, Methodology, Software, Investigation, Data Curation, Visualization, Writing - Original Draft, Writing - Review \& Editing. {\bf Haroldo V. Ribeiro:} Conceptualization, Methodology, Software, Investigation, Data Curation, Visualization, Writing - Original Draft, Writing - Review \& Editing.

\section*{Declaration of competing interest}
\noindent The authors declare that they have no known competing financial interests or personal relationships that could have appeared to influence the work reported in this paper.

\begin{acknowledgments}
We acknowledge the support of the Coordena\c{c}\~ao de Aperfei\c{c}oamento de Pessoal de N\'ivel Superior (CAPES -- PROCAD-SPCF Grant 88881.516220/2020-01) and the Conselho Nacional de Desenvolvimento Cient\'ifico e Tecnol\'ogico (CNPq -- Grants 303533/2021-8 and 309560/2025-0), and Slovenian Research and Innovation Agency (Javna agencija za znanstvenoraziskovalno in inovacijsko dejavnost Republike Slovenije) (Grant P1-0403). We also thank Prof. Cristina Masoller for useful discussions during the ENFE 2024.
\end{acknowledgments}

\section*{Data availability}
The data used in the current study are available from Refs.~\cite{ribeiro2018thedynamical, martins2022universality} and the corresponding author upon reasonable request.

\bibliography{references.bib}

\end{document}